\documentclass[a4paper]{jpconf}
\usepackage{graphicx}
\usepackage{epstopdf}

\begin{document}

\title{Very-high energy emission from pulsars}

\author{M Breed$^1$, C Venter$^1$ and A K Harding$^2$}

\address{$^1$ Centre for Space Research, North-West University, Potchefstroom Campus, Private Bag X6001, Potchefstroom, 2520, South Africa}
\address{$^2$ Astrophysics Science Division, NASA Goddard Space Flight Center, Greenbelt, MD 20771, USA}

\ead{20574266@nwu.co.za}

\begin{abstract}
The vast majority of pulsars detected by the {\it Fermi} Large Area Telescope (LAT) display exponentially cutoff spectra with cutoffs falling in a narrow band around a few GeV. Early spectral modelling predicted spectral cutoffs at energies of up to 100 GeV, assuming curvature radiation. It was therefore not expected that pulsars would be visible in the very-high energy (VHE) regime ($>100$ GeV). The \textit{VERITAS} announcement of the detection of pulsed emission from the Crab pulsar at energies up to $400$ GeV (and now up to 1.5~TeV as detected by \textit{MAGIC}) therefore raised important questions about our understanding of the electrodynamics and local environment of pulsars. \textit{H.E.S.S.} has now detected pulsed emission from the Vela pulsar down to tens of GeV, making this the second pulsar detected by a ground-based Cherenkov telescope. Deep upper limits have also been obtained by \textit{VERITAS} and \textit{MAGIC} for the Geminga pulsar. We will review the latest developments in VHE pulsar science, including an overview of the latest observations, refinements, and extensions to radiation models and magnetic field structures, and the implementation of new radiation mechanisms. This will assist us in understanding the VHE emission detected from the Crab pulsar, and predicting the level of VHE emission expected from other pulsars, which is very important for the upcoming \textit{CTA}.
\end{abstract}

\section{Introduction}

Since the launch in June 2008 of {\it Fermi} LAT \cite{Atwood2009}, a high-energy (HE) satellite measuring $\gamma$-rays in the range 20 MeV$-$300 GeV, two pulsar catalogues (1PC, \cite{Abdo2010}; 2PC, \cite{Abdo2013}) discussing the light curve and spectral properties of 117 pulsars have been released. The vast majority of the {\it Fermi}-detected pulsars display exponentially cutoff spectra with cutoffs around a few GeV. These spectra are believed to be due to curvature radiation (CR), which is assumed to be the dominating emission process in the GeV band (see Section~\ref{subsection:radiationmechanisms}). 

\subsection{Standard pulsar emission models}\label{subsection:emissionmodels}

There exist several physical radiation models that can be used to study HE emission from pulsars. These include the polar cap (PC; \cite{Daugherty1982}), slot gap (SG; \cite{Arons1983}), outer gap (OG; \cite{Cheng1986b}), and the pair-starved polar cap (PSPC; \cite{Harding2005}) models, which can be distinguished from each other based on the different assumptions of the geometry and location of the `gap regions'. The `gap region' is where particle acceleration takes place due to an unscreened, rotation-induced $E$-field parallel to the local $B$-field, as well as subsequent emission by these particles. 

In PC models emission from HE particles is assumed to originate close to the neutron star (NS) surface. These particles are accelerated by large $E$-fields near the magnetic poles (known as the magnetic PCs) only up to a few stellar radii. In SG models, the radiation comes from narrow gaps close to the last open field lines (the field lines that are tangent to the light cylinder where the corotation speed equals the speed of light $c$), with the gaps extending from the NS surface up to high altitudes. In the OG model, the gap region extends from the null-charge surface, where the Goldreich-Julian charge density is zero \cite{Goldreich1969} up to high altitudes, also close to the last open field lines. The PSPC model involves a gap region that extends from the NS surface to the light cylinder over the full open volume \cite{Harding2005}, since the potential is unscreened in this case, so that there are not enough pairs to fully screen the $E$-field.

\subsection{Radiation and pair creation processes}\label{subsection:radiationmechanisms}

To explain HE emission in the standard models, one has to take detailed particle transport and radiation mechanisms into account. These mechanisms include CR, synchrotron radiation (SR), and inverse Compton scattering (ICS). CR occurs whenever charged particles are constrained to move along curved paths, e.g., along curved $B$-field lines (e.g., \cite{Harding1981}), therefore involving a change in their longitudinal kinetic energy. When the emitted CR photon energy and the local $B$-field are high enough, magnetic pair production may occur (where an HE photon converts into an electron-positron pair, $e^\pm$), leading to a cascade of $e^\pm$ pairs which may screen the parallel $E$-field outside the gaps. The pair cascade is characterized by the so-called multiplicity, i.e., the number of pairs spawned by a single primary. The pairs may radiate SR if they have velocity components perpendicular to the local $B$-field so that this process involves a change in the particles' transverse kinetic energy. Also, ICS occurs due to the relativistic particles which upscatter soft photons (e.g., originating at a heated PC), which results in the ``boosting'' of the photon energies up to very high energies. ICS photons may also be converted into $e^\pm$ pairs. Two-photon pair creation ($\gamma\gamma$-absorption) may also occur, in particular in OG models. Cyclotron emission combined with subsequent ICS has also been considered by \cite{Lyutikov2013} to explain the broadband spectrum of the Crab pulsar.

\subsection{Historic perspective of VHE spectral modelling}

\begin{figure}[t]\centering
	\includegraphics[width=22.3pc,height=13pc]{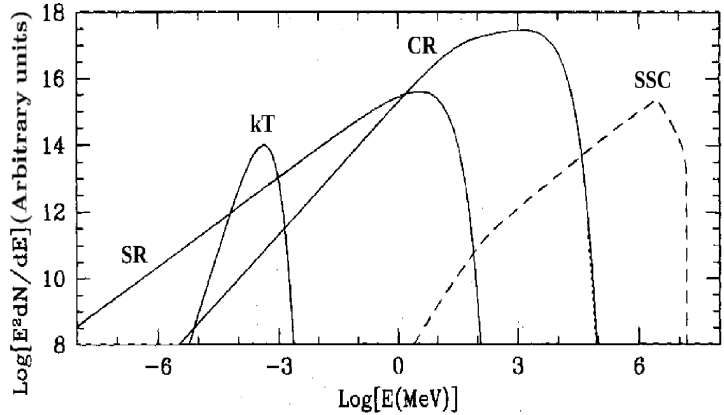}\hspace{0.5cm}
\caption{\label{Romani96} Early prediction of the phase-averaged spectrum for the Vela pulsar. The solid lines represent the spectral components for CR, SR, and the thermal surface flux (kT). The dashed curve represents the TeV pulsed spectral component associated with the ICS of SR of the primary $e^\pm$ (SSC). Adapted from \cite{Romani1996}.}
\end{figure}
\begin{figure}[t]\centering
	\includegraphics[width=20pc,height=15pc]{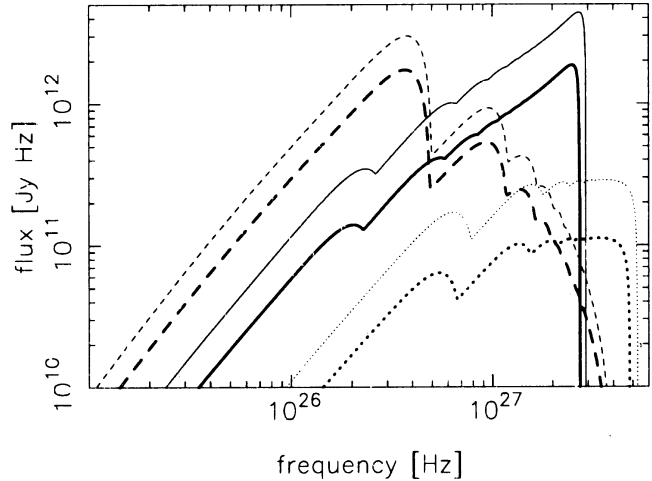}\hspace{0.5cm}
\caption{\label{Hirotani01} Expected TeV spectra for three bright pulsars including the Crab (solid lines), PSR B0656$+$14 (dashed lines), and PSR B1509$-$58 (dotted lines). The thick and thin curves represent inclination angles (between the spin and magnetic axis) of $\alpha=30^\circ$ and $45^\circ$, respectively \cite{Hirotani2001a}.}
\end{figure}

Early modelling, assuming the standard OG model, predicted spectral components in the VHE regime when estimating the ICS of primary electrons on SR or soft photons. This resulted in a natural bump around a few TeV (involving $\sim 10$ TeV particles) in the extreme Klein-Nishina limit as seen in figure~\ref{Romani96} and \ref{Hirotani01}. However, these components may not survive up to the light cylinder and beyond \cite{Cheng1986b,Romani1996,Hirotani2001a}, since $\gamma\gamma$ pair creation leads to absorption of the TeV $\gamma$-ray flux.
Other studies assumed CR to be the dominant radiation mechanism producing $\gamma$-ray emission when performing spectral modelling and found spectral cutoffs of up to 50 GeV. For example, \cite{Bulik2000} modelled the cutoffs of millisecond pulsars (MSPs). These are pulsars which possess relatively low $B$-fields and short periods. Their model assumed a static dipole $B$-field and a PC geometry, and predicted CR from the primary electrons that are released from the PC. Their predicted CR spectral component cut off at $\sim100$~GeV. The CR photons may undergo pair production in the intense low-altitude $B$-fields, and the newly formed electron-positron secondaries will emit SR in the optical and X-ray band (see \cite{Harding2002}). Therefore, they concluded that the HE CR from MSPs occurred in an energy band that was above the detection range of satellite detectors like \textit{Energetic Gamma-Ray Experiment Telescope (EGRET)} and below that of ground-based Cherenkov detectors such as the \textit{High Energy Stereoscopic System (H.E.S.S. I)}.

Later studies investigated the X-ray and $\gamma$-ray spectrum of rotation-powered MSPs using a PSPC model \cite{Harding2005}, and found CR cutoffs of $\sim10-50$~GeV (see \cite{Frackowiak2005,Venter2005}). Optical to $\gamma$-ray spectra were also modelled by \cite{Harding2008} assuming an SG accelerator and a retarded vacuum dipole (RVD) $B$-field, for the Crab pulsar. They found spectral cutoffs of up to a few GeV. Another study modelled the phase-resolved spectra of the Crab pulsar using the OG and SG models, and found HE cutoffs of up to $\sim25$~GeV \cite{Hirotani2008}. Cutoffs around $\sim10$~GeV were found for the OG model using the RVD $B$-field \cite{Tang2008}. 

\section{Observational revolution}

In view of the above theoretical paradigm it was not expected that a pulsar should be visible in the VHE regime. It was therefore surprising when the \textit{Very Energetic Radiation Imaging Telescope Array System (VERITAS)} announced the detection of pulsed emission from the Crab pulsar above $\sim100$~GeV \cite{Aliu2011}, followed by the detection by the \textit{Major Atmospheric Gamma-Ray Imaging Cherenkov (MAGIC)} of emission up to $\sim400$~GeV (soon after their initial detection of emission at $\sim25$~GeV) \cite{Aleksic2012,Aliu2008}. The \textit{MAGIC} Collaboration has since reported the detection of pulsed photons with energies up to 1.5~TeV \cite{Ansoldi2016}. Ground-based Cherenkov telescopes are now searching for more examples of VHE pulsars.
\begin{figure}[t]\centering
	\includegraphics[width=18pc,height=18pc]{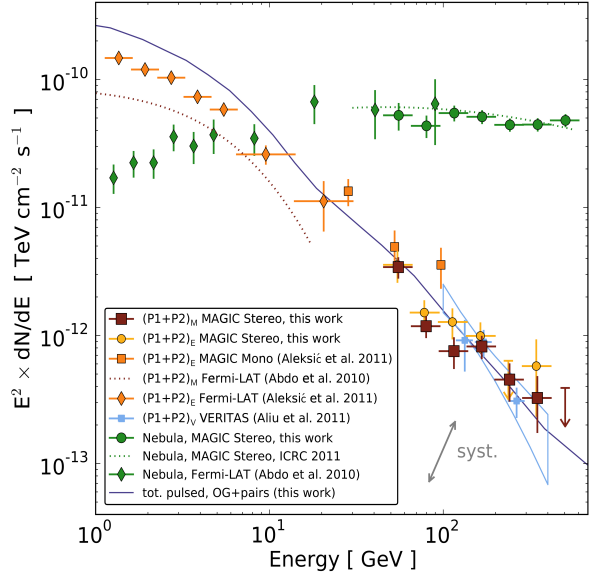}\hspace{0.5cm}
\caption{\label{MAGIC} The observed and modelled spectra for phase-resolved $\gamma$-ray emission, including both emission peaks P1 and P2, from the Crab pulsar as measured by \textit{MAGIC} (dark red squares). The plot also contains measurements from \textit{Fermi} LAT (orange diamonds) and \textit{VERITAS} (light blue squares and solid line). An OG model including emission from pairs are assumed. The systematic error of the \textit{MAGIC}-Stereo measurement is shown and corresponds to a shift of $\pm$17\% in energy and $\pm$19\% in flux. From \cite{Aleksic2012}.}
\end{figure}

In figure~\ref{MAGIC} the phase-resolved $\gamma$-ray spectrum for the Crab pulsar for both light curve peaks P1 and P2, as measured by \textit{MAGIC}, is shown. The spectrum also contains measurements by \textit{Fermi} LAT and \textit{VERITAS}. The observations by \textit{MAGIC} are in good agreement with those by \textit{VERITAS}. The \textit{VERITAS} data fit a broken power law, and raised important questions whether the CR component is extended or if a second component is required to explain the observed spectra.

The detection of the Crab pulsar above several GeV prompted {\it Fermi} to search for pulsed emission at HEs. They detected significant pulsations above 10~GeV from 20 pulsars and above 25~GeV from 12 pulsars \cite{Ackermann2013}. The Crab pulsar is the first source which have been detected over almost all energies ranging from radio to VHE $\gamma$-rays. More recently, pulsed emission was detected from the Vela pulsar above 30 GeV with the \textit{H.E.S.S.} \cite{Stegmann2014} and up to $\sim80$~GeV (at the $4\sigma$-level) with the \textit{Fermi} LAT \cite{Leung2014}. \textit{MAGIC} furthermore detected no emission from Geminga above 50 GeV~\cite{Ahnen2016}; neither did \textit{VERITAS} above 100~GeV~\cite{Aliu2015}. A stacking analysis involving 115 {\it Fermi}-detected pulsars (excluding the Crab pulsar) was performed by \cite{McCann2015} using {\it Fermi} data. However no emission above 50 GeV was detected, implying that VHE pulsar detections may be rare, given current telescope sensitivities. From all these observations there are three effects visible in the energy-dependent pulse profiles: the peaks remain at the same phase, the P1/P2 ratio decreases as energy increases, and the pulse width decreases with increasing energy. More VHE pulsars may be found by the \textit{Cherenkov Telescope Array (CTA)} which will have a ten-fold increase in sensitivity compared to present-day Cherenkov telescopes.

\section{Theoretical ideas}

All of the standard pulsar emission models (see Section~\ref{subsection:emissionmodels}) predicted HE spectral cutoffs between a few GeV and up to $\sim100$~GeV, assuming $B$-fields such as the static dipole and RVD solutions. Clearly, refinements to these radiation models and $B$-fields are needed to explain the observed VHE emission from the Crab. There are a few ideas for such refinements.  One is a revised OG model by \cite{Hirotani2008} which can produce IC radiation of up to $\sim400$~GeV due to secondary and tertiary pairs upscattering infrared to ultraviolet photons \cite{Aleksic2012}. In this OG model the IC flux depends sensitively on the $B$-field structure near the light cylinder. 

\begin{figure}[t]\centering
	\includegraphics[width=22.3pc]{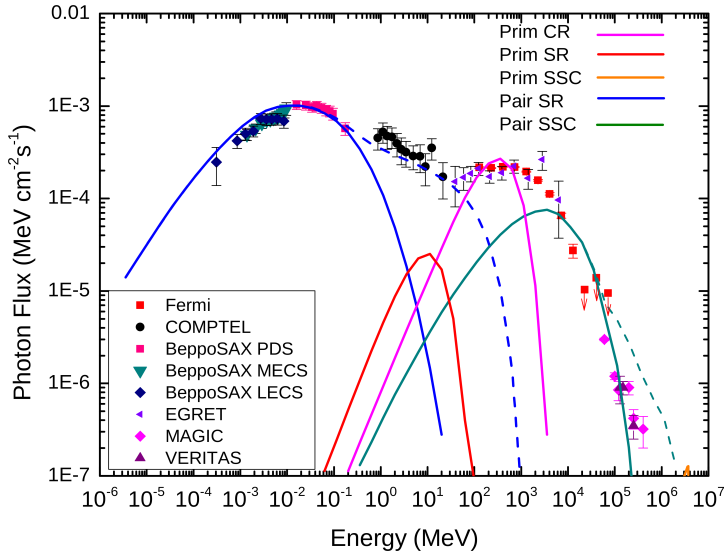}\hspace{0.5cm}
\caption{\label{SSC} Modelled spectrum of phase-averaged pulsed emission from the Crab pulsar. The spectral components from primary electrons and pairs (as labelled) are for a magnetic inclination angle $\alpha=45^\circ$, observer angle $\zeta= 60^\circ$ and pair multiplicity $M_{+} = 3\times10^5$. From \cite{Harding2015}.}
\end{figure}

Another idea was proposed by \cite{Lyutikov2012}, invoking the SSC radiation process. This is indeed a promising radiation mechanism, where relativistic particles upscatter the SR photons emitted by the same population. The SSC radiation mechanism was applied by \cite{Harding2015} to predict optical to X-ray, and $\gamma$-ray spectra (see figure~\ref{SSC}) assuming an SG model and a force-free $B$-field. This process relies critically on the assumed electrodynamics and the magnetospheric structure. They performed simulations for the Crab and Vela pulsars, as well as two MSPs, i.e., B1821$-$24 and B1937$+$21. However, the only significant predicted SSC component was for the Crab pulsar. They also found that the pair SR matched the observed X-ray spectrum of the MSPs. They furthermore tested the addition of an HE power law extension to the pair spectrum (dashed lines in figure 4) whose SR spectrum would account for the observed emission in the $1-100$~MeV range. However, the resulting SSC component exceeded the observed \textit{MAGIC} and \textit{VERITAS} points, implying that the observed $1-100$~MeV emission is not produced by the same particles that produce the SSC emission.

\section{Conclusions and future prospects}

The abovementioned detections of the VHE pulsed emission from pulsars and the explanation thereof implies that this emission may yield strong constraints on $\gamma$-ray radiation mechanisms, the location of acceleration regions, and the $B$-field structure. There is thus an urgent need for refinements and extensions of standard pulsar emission models and radiation mechanisms, including more realistic $B$-fields. Some examples are discussed above, but there are many more. 

SG model refinements include photon-photon pair production attenuation within the model and also more realistic $B$-fields such as dissipative magnetospheric solutions. One could also model the emission pulse profiles as a function of energy.

Ground-based Cherenkov telescopes are now searching for more examples of VHE pulsars. New pulsar models will assist us in predicting the level of VHE emission expected from them, which would be very important for the upcoming \textit{CTA}. The low threshold energy of \textit{CTA} will provide an overlap with the {\it Fermi} energy range and will help to discriminate between CR and a potentially new spectral component.

\ack
This work is based on research supported by the National Research Foundation (NRF) of South Africa (Grant Numbers 90822, 93278, and 99072). Any opinions, findings, and conclusions or recommendations expressed are that of the authors, and the NRF accepts no liability whatsoever in this regard. A.K.H.\ acknowledges the support from the NASA Astrophysics Theory Program.

\end{document}